\documentclass[aps,pra,showpacs,notitlepage]{revtex4-1}

\usepackage{color,graphicx}
\begin{document}

%%\title{How much physics is there in quantum statistics?}

\title{What does the operator algebra of quantum statistics tell us\\
about the objective causes of observable effects?}

\author{Holger F. Hofmann}
\email{hofmann@hiroshima-u.ac.jp}
\affiliation{
Graduate School of Advanced Science and Engineering, Hiroshima University,
Kagamiyama 1-3-1, Higashi Hiroshima 739-8530, Japan
}

\begin{abstract}
Quantum physics can only make statistical predictions about possible measurement outcomes, and these predictions originate from an operator algebra that is fundamentally different from the conventional definition of probability as a subjective lack of information regarding the physical reality of the system. In the present paper, I explore how the operator formalism accommodates the vast number of possible states and measurements by characterizing its essential function as a description of causality relations between initial conditions and subsequent observations. It is shown that any complete description of causality must involve non-positive statistical elements that cannot be associated with any directly observable effects. The necessity of non-positive elements is demonstrated by the uniquely defined mathematical description of ideal correlations which explains the physics of maximally entangled states, quantum teleportation and quantum cloning. The operator formalism thus modifies the concept of causality by providing a universally valid description of deterministic relations between initial states and subsequent observations that cannot be expressed in terms of directly observable measurement outcomes. Instead, the identifiable elements of causality are necessarily non-positive and hence unobservable. The validity of the operator algebra therefore indicates that a consistent explanation of the various uncertainty limited phenomena associated with physical objects is only possible if we learn to accept the fact that the elements of causality cannot be reconciled with a continuation of observable reality in the physical object.
\end{abstract}

\maketitle
%%--Introduction

\section{Introduction}
Perhaps the greatest puzzle of quantum theory is the fact that it does not provide us with any feasible model of physical reality, even though it allows us to describe a vast number of possible experimental scenarios and does provide us with accurate predictions of the statistics obtained from their outcomes. How can it be that all of this evidence does not provide us with a clearer picture of the physics represented by the theory? This question seems to be all the more urgent given the emergence of a whole new class of quantum technologies in the wake of quantum information and quantum computation related research. It should not be forgotten that an important motivation of many researchers in this new field of physics has been the development of a better understanding of quantum mechanics based on the analysis of quantum processes as a form of information processing \cite{Zei99,Bru99,Fuc03,Cav07,Goy08,Lee11,Lei13}. From the experimental side, the goal of establishing a more complete control over the non-classical properties of quantum systems has made it necessary to introduce new methods of characterizing the precision achieved by these efforts, resulting in the development of complete quantum state and quantum process tomographies \cite{Leo96,Whi99,Res05,Rie06}. This kind of complete characterization of quantum statistics is a direct application of the operator algebra of quantum mechanics to the new experimental possibilities of emerging quantum technologies \cite{Mahler}. It has always astonished me how little impact this practical demonstration of the power of operator algebra has had on our fundamental understanding of quantum physics. The problem seems to be that even the experimentalists in the field take it for granted that the formalism defines a reality that is separate from the practical ``shadow'' it casts in each individual experiment. Unfortunately, there is some truth to this idea, since experiments can never be universal enough to reveal the underlying physics without a theoretical hypothesis.
As a result, much of the work in the quantum information community has focused on attempts to reconstruct causality relations for specific protocols characterized by binary choices \cite{Lee11,Lei13}. However, such information theoretic approaches ignore the consistency of the Hilbert space description and the obvious relations between incompatible measurements. Explanations of causality should work equally well for all possible phenomena and a proper formulation of the theory must focus on the connection between observable phenomena and the universal mathematical structure that provides a consistent description despite the physical differences between the different scenarios.

In classical physics, the tight connection between phenomena and theory is achieved by the identification of quantitative observations with real numbers that represent the coordinates of a geometric manifold, where different observations can be identified with specific regions on the manifold. The problems with quantum physics arise because the statistical predictions of the formalism cannot be traced back to a unique set of points representing individual realities. 
Instead, the quantum formalism provides an alternative representation of statistics that appears to change from situation to situation, making a non-contextual description of measurement outcomes impossible. This irreducible contextuality of quantum statistics has been enshrined in the form of the uncertainty principle, where it describes the relation between physical properties that have no observable joint reality. The problem with the uncertainty principle is that it is not really a principle at all. Instead, it represents only one aspect of a formalism that gives a very specific description of the relation between different states and measurements. Consequently, there are many ways to address the problem of uncertainty within the framework of the operator formalism \cite{Oza03,Hof03,Wat11,Bus13,Dre14} and none of these approaches have been able to clarify the nature and role of uncertainty in quantum physics in a satisfactory manner. For a more fundamental explanation of the mysterious absence of elementary realities, it may therefore be more useful to turn to the formalism itself in order to  identify the fundamental mathematical features that make a proper explanation of quantum phenomena so difficult. In particular, we should explain in detail how the operator formalism establishes the essential connection between mathematical expressions and experimentally observable phenomena in the wide range of possible experimental scenarios. Since the physical meaning of the formalism has to be defined in terms of the experimentally observable phenomena it describes, the formalism can only be understood in terms of its contributions to experimentally testable results.

In the following, I will use a particularly compact formulation of quantum statistics to investigate the universal characteristics of all complete representations of quantum statistics as expressed by the operator algebra. As will be shown below, these characteristics do not satisfy the requirements of a conventional statistical interpretation in terms of a manifold or set of uniquely defined ``elements of reality.'' Instead, the formalism defines ``elements of causality'' that provide an alternative description of perfectly deterministic relations between state preparation and measurement.  These elements are fundamentally unobservable because they represent non-positive contributions to the statistics. The positivity of actual experimental results can only be ensured by the constraints imposed on state preparation and measurement associated with quantum uncertainties. Because of this uncertainty constraint, the experimentally accessible reality of a system cannot provide a complete description of causality in terms of directly observable effects. However, it is possible to find experimentally observable phenomena that are characterized by ideal correlations between quantum systems. Specifically, ideal correlations appear in the characterization of maximally entangled states and of quantum teleportation. In addition, optimal quantum cloning is characterized by a well-defined contribution that represents an ideal copy of an arbitrary input. All of these phenomena support the idea that complete operator expansions represent  a  universal description of causality in quantum physics, with the elements of any complete and orthogonal expansion serving as deterministic ``elements of causality.'' The operator formalism itself thus provides a description of causality that can replace the redundant hypothesis of a continuation of observable reality inside the physical object. Instead, the universal causality relations that explain all of the possible phenomena associated with a physical object are necessarily expressed in terms of non-positive elements that need to be understood in relation to the uncertainty limits that apply to the possible external controls by means of which the physical object is manipulated and observed. 

\section{A compact theory of quantum statistics}
\label{sec:compact}

Quantum theory is usually formulated in terms of quantum states represented by Hilbert space vectors $\mid \psi \rangle$ or the corresponding density operators $\hat{\rho}$. It is assumed that these states describe specific physical situations. However, the uncertainty principle strictly limits the amount of control that the initial conditions of quantum state preparation can give us over the actual physical properties of an individual system. In the mathematical formalism, the uncertainty principle is included in Hilbert space as a fundamental relation between the outcomes of projective measurements. It is therefore impossible to explain the physics of quantum states without somehow referring to the outcomes of quantum measurements. 

If a quantum measurement is formulated in terms of its possible outcomes $\{ b \}$, its Hilbert space representation is given by a positive operator-valued measure (POVM) $\{ \hat{E}(b) \}$ representing the probabilities of each outcome $b$. Specifically, the probability to obtain $b$ from an initial condition $a$ represented by a density operator $\hat{\rho}(a)$ is given by the product trace of the two operators representing the initial conditions and the outcome,
\begin{equation}
\label{eq:prob}
P(b|a) = \mbox{Tr}\left(\hat{E}(b) \hat{\rho}(a) \right). 
\end{equation}
This formulation relates any possible initial condition $a$ to any possible observation $b$ and therefore summarizes all of the physics described by the Hilbert space formalism in a generalized description of causality valid for all possible quantum systems. Significantly, this description of causality cannot be reduced to more fundamental individual realities $i$, and it may be worth considering what aspects of the mathematical structure prevent such a more intuitive explanation of the statistics observed in quantum systems. 

Essentially, the question can be formalized by considering an expansion of the product trace in Eq.(\ref{eq:prob}). Using the terminology of probability theory, we can first decompose $\hat{\rho}(a)$ into statistical coefficients $P(i|a)$ associated with a set of elements $i$ and then define the measurement probabilities contributed by each element $i$ as $P(b|i)$. The inner product in Eq.(\ref{eq:prob}) then takes on the form familiar from conventional statistics,
\begin{equation}
\label{eq:elements}
P(b|a) = \sum_i P(b|i) P(i|a). 
\end{equation}
Mathematically, this is merely a different representation of the same inner product. It is therefore important to consider the possible representations $i$ of the product trace and identify its characteristics for positive self-adjoint operators $\hat{\rho}(a)$ and $\hat{E}(b)$. Operator expansions of this form have been known for a very long time, and much has been written about them \cite{Mahler}. However, it seems to me that their mathematical simplicity has distracted researchers from the oddities of the physics described in this manner. The critical point is the association of each element $i$ with a specific set of operators,
\begin{eqnarray}
P(i|a) &=& \mbox{Tr}\left(\hat{\Lambda}(i) \hat{\rho}(a) \right),
\nonumber \\ 
P(b|i) &=& \mbox{Tr}\left(\hat{E}(b) \hat{R}(i) \right). 
\end{eqnarray}
It is easy to see that the representation of the inner product in Eq.(\ref{eq:elements}) then requires that 
\begin{eqnarray}
\label{eq:ortho}
\mbox{Tr}\left(\hat{\Lambda}(i) \hat{R}(j) \right) = \delta_{i,j}.
\end{eqnarray}
Specifically, $\hat{\Lambda}(i)$ identifies the contribution of $i$ to any measurement outcome $\hat{E}(b)$, while $\hat{R}(i)$ identifies the statistical weight of $i$ in the initial conditions $\hat{\rho(a)}$. 

If Hilbert space merely encoded classical statistics, it would be possible to identify a unique set of realities $i$ that could then be used to describe the deterministic laws of physics governing the causality relations of the quantum system. However, the operator expansion associated with such a set of realities would have to satisfy three conditions,

\begin{enumerate}
\item 
{\bf Positivity:} 
\\
It should be possible to realize a measurement of the elements $i$, and this measurement must be represented by $\hat{\Lambda}(i)$. Therefore, the operators $\hat{\Lambda}(i)$ should be self-adjoint and positive. 

\item
{\bf Orthogonality:}
\\
It should be possible to distinguish the elements from each other. This is only possible if statistical elements have no overlap with each other. According to Eq.(\ref{eq:ortho}), this condition is satisfied if $\hat{\Lambda}(i)=\lambda_i \hat{R}^\dagger(i)$, where $\lambda_i$ represents the difference in the normalization of $\hat{\Lambda}(i)$ and $\hat{R}(i)$.

\item
{\bf Completeness:}
\\
It should be possible to explain any causality relation between initial conditions and measurement outcomes in terms of the complete set of elements $\{i\}$. Therefore, it should be possible to express any quantum state by a linear combination of the elements $\{\hat{R}(i)\}$ and any measurement by linear combinations of the elements $\{\hat{\Lambda}(i)\}$.  
\end{enumerate}

As I will discuss in more detail in the following, the adjoint sets of operators $\{\hat{\Lambda}(i)\}$ and $\{\hat{R}(i)\}$ can never satisfy more than two of the three requirements listed above. It is therefore impossible to apply the conventional interpretation of probability as a representation of subjective ignorance to quantum statistics. Instead, more thought needs to be given to the three individual requirements shown above and the experimental situations to which they apply. 

\section{Precise measurements}
\label{sec:precise}
 
Sets of operators that satisfy both condition 1 and condition 2 represent possible measurements with distinguishable outcomes. The only POVMs that have no overlap between different elements are POVMs that consist entirely of projection operators with eigenvalues of zero or one. The maximal set of operators satisfying conditions 1 and 2 is therefore given by a sets of projectors onto orthogonal basis states ${\mid n \rangle}$ that represent a precise measurement of a specific physical property of the system,
\begin{equation}
\label{eq:project}
\hat{\Lambda}(n) = \mid n \rangle \langle n \mid. 
\end{equation}
Although this POVM describes a precise measurement, the information obtained in the measurement is incomplete. Specifically, the operators $\hat{R}(n)=\hat{\Lambda}(n)$ representing the corresponding quantum state preparations do not describe any coherences between different basis states $\mid n \rangle$. Thus, the operator expansion is incomplete and does not satisfy condition 3. On the other hand, condition 1 is satisfied because the outcomes $n$ represent valid measurement results, and condition 2 is satisfied because the measurement results distinguish different states of the system. Thus conditions 1 and 2 define a precise measurement of a specific property of the system, without any disturbance from external noise sources. 

It is worth noting that the original formulation of quantum mechanics seems to imply that precise measurements satisfying conditions 1 and 2 represent the only valid statements one can make about a quantum system. The problem that such measurements are necessarily incomplete is usually associated with the uncertainties of quantum states, introducing an asymmetry in the role of states and of measurements that is not quite justified by the formalism itself. This kind of unwarranted bias in the interpretation of quantum states overlooks the important role of the causality relations expressed by the product traces in Eq. (\ref{eq:prob}), which can only be explained in a consistent manner if the same analysis is used to characterize both measurements and states. It is therefore necessary to introduce complete sets of intermediate elements $i$ to describe the causality relation between state and measurement outcome. In the conventional approach, these elements are added in an ad hoc manner by introducing the off-diagonal elements of the density matrix in the $\{\mid n \rangle\}$-basis, represented by additional elements of the operator expansion
\begin{equation}
\label{eq:complete}
\hat{\Lambda}(n,n^\prime) = \mid n \rangle \langle n^\prime \mid,
\end{equation}
where $\hat{R}(n,n^\prime)=\hat{\Lambda}^\dagger(n,n^\prime)$. This expansion satisfies conditions 2 and 3, but fails to satisfy condition 1 for all $n\neq n^\prime$. This strange inconsistency between diagonal elements and off-diagonal elements in the conventional density matrix description of quantum states prevents a straightforward interpretation of the formalism in terms of specific measurement outcomes. It is therefore useful to consider other ways to obtain a complete description of quantum statistics, without the artificial separation of elements into directly observable probabilities and quantum coherences that can only be reconstructed from the outcomes of completely different measurements. 

In the following discussion, the goal is to discuss the physics associated with complete representations of the operator formalism. The density matrix expansion is not ideal for this purpose because it is not useful to enforce condition 1 for only some of the elements in the operator expansion, while accepting negative and even complex values for the all of the others. Although the density matrix representation is complete and orthogonal, it is artificially biased in favor of the basis $\mid n \rangle$, obscuring the perfectly equivalent physics of other possible basis states. In this context, it seems to make more sense to consider uncertainty limited joint measurements of the non-commuting physical properties of the system. In particular, such measurements can satisfy conditions 1 and 3, providing complete information about the state of the system without biases in favor of specific physical properties.

\section{Tomographic reconstructions}

Any operator expansion that satisfies condition 3 is a complete representation of all possible operators of that Hilbert space. The representation of any density operator $\hat{\rho}(a)$ is given by the expansion
\begin{equation}
\label{eq:statesum}
\hat{\rho}(a)=\sum_i \mbox{Tr}(\hat{\Lambda(i)}\hat{\rho}(a)) \; \hat{R}(i).
\end{equation}
If the operators $\hat{\Lambda}(i)$ satisfy condition 1, it is possible to reconstruct the quantum state $\hat{\rho}(a)$ from the experimentally observable probabilities $P(i|a)$ obtained from measurements of $i$,
\begin{equation}
\hat{\rho}(a)=\sum_i P(i|a) \; \hat{R}(i).
\end{equation}
In general, the measurement outcome $i$ partially depends on external noise. The reconstruction operators $\hat{R}(i)$ represent a deconvolution of this measurement noise. In the operator formalism, the presence of measurement errors is represented by the overlap between the operators $\hat{\Lambda}(i)$ representing different measurement outcomes $i$. Since the deconvolution of noise results in negative probabilities when applied to a noise-free distribution, the presence of noise in the POVM $\{\hat{\Lambda}(i)\}$ appears in the form of negative eigenvalues in the reconstruction operators $\hat{R}(i)$. Any tomographic reconstruction of quantum states therefore involves non-positive elements $\hat{R}(i)$, and the negativity of the operators  $\hat{R}(i)$ represents the amount of background noise in the measurement represented by the POVM $\{\hat{\Lambda}(i)\}$.

To quantify the necessary amount of background noise in the measurement probability $P(i|a)$, it is useful to consider a peculiar mathematical property of any complete operator expansion that can give a specific quantitative meaning to the impossibility of satisfying conditions 1 and 2. The mathematical property in question can be derived directly from the expansion given in Eq.(\ref{eq:statesum}). By applying this expansion to the product trace that defines the probability of a measurement outcome $b$ represented by $\hat{E}(b)$, we find the expression
\begin{equation}
\mbox{Tr}\left(\hat{E}(b) \hat{\rho}(a) \right) = \sum_i \mbox{Tr}\left(\hat{E}(b) \hat{R}(i) \right) \; \mbox{Tr}\left(\hat{\Lambda}(i) \hat{\rho}(a) \right).
\end{equation}
The expansion of a single product trace therefore separates the product traces of $\hat{\rho}(a)$ from the product traces of $\hat{E}(b)$. We can use this separation to define two separate Hilbert spaces, one for the state expansion and one for the expansion of the measurement. It is then possible to write the product trace within one Hilbert space as a product trace of two Hilbert spaces, where the expansion into elements $i$ is represented by a sum over the dyadic products of the basis operators of the expansion,
\begin{equation}
\mbox{Tr}\left(\hat{E}(b) \hat{\rho}(a) \right) = \mbox{Tr}\left(
\left(\hat{E}(b) \otimes \hat{\rho}(a)\right) \; \left(\sum_i \hat{R}(i) \otimes \hat{\Lambda}(i)\right)\right).
\end{equation}
This relation uniquely defines the operator expressed by the sum over $i$. The only operator for which the product trace of a dyadic product of operators with that operator is equal to the product trace of the two operators in the dyadic product is the operator $\hat{U}_{\mathrm{SWAP}}$ describing an exchange of the two Hilbert spaces, with its eigenvalues of $+1$ for all states that are symmetric under the exchange and of $-1$ for all states that are anti-symmetric under the exchange. 
\begin{equation}
\label{eq:swap}
\sum_i \hat{R}(i) \otimes \hat{\Lambda}(i) = \hat{U}_{\mathrm{SWAP}}.
\end{equation}
This relation applies to all complete expansions of the operator algebra, irrespective of the specific choice of the operators $\hat{\Lambda}_i$ and $\hat{R}_i$. Eq.(\ref{eq:swap}) thus expresses a necessary mathematical property of all complete operator expansions.
Since the operator $\hat{U}_{\mathrm{SWAP}}$ has negative eigenvalues, Eq.(\ref{eq:swap}) shows that all complete operator expansions involve non-positive operators, either in  $\{\hat{\Lambda}(i)\}$, or in $\{\hat{R}(i)\}$, or in both. It is therefore impossible to find an operator expansion that is both complete and positive. 

For operator expansions satisfying condition 1, the operators $\hat{\Lambda}(i)$ are positive and represent a noisy measurement with an amount of noise reflected by the negativity of the reconstruction operators $\{\hat{R}(i)\}$. A good estimate of the minimal amount of negativity in the reconstruction operators $\{\hat{R}(i)\}$ can be obtained by considering the addition of multiples of the identity operator $\hat{I}$ to the reconstruction operators $\hat{R}(i)$. Since $\{\hat{\Lambda}(i)\}$ is a POVM, a positive result is obtained with
\begin{equation}
\label{eq:fill}
\sum_i \left(\hat{R}(i)+\hat{I}\right)\otimes \hat{\Lambda}(i) = \hat{U}_{\mathrm{SWAP}} + \hat{I} \otimes \hat{I}.
\end{equation}
The operator on the right hand side is the projector onto the symmetric states, multiplied by a factor of two. The addition of identity operators to each reconstruction operator is therefore the minimal addition needed to compensate the negative eigenvalues of the operators $\hat{R}(i)$. Given that the traces of the operators are one, the necessary presence of negative eigenvalues of $-1$ in the reconstruction operators $\hat{R}(i)$ indicated by Eq.(\ref{eq:fill}) shows that a large amount of overlap between the elements of the POVM $\{\hat{\Lambda}(i)\}$ is necessary to achieve tomographic completeness. It is therefore difficult to identify the physical meaning of each measurement outcome $i$ within the system.

\section{Statistical moments and quasi-probabilities}

Experimentally observable statistics are either noisy or incomplete. Both problems make it difficult to identify the precise causality relations within the system. It may therefore be reasonable to consider non-positive statistical characterizations instead, satisfying conditions 2 and 3 to provide a complete and distinguishable set of orthogonal elements to characterize the physics of the system. The prime example of such an approach is the density matrix representation discussed above. In that description, quantum coherences can be interpreted as statistical moments that appear as quantum interference patterns in the probability distributions of other physical properties of the system. The most striking example is the identification of the modulation of probability in the double slit interference experiment with off-diagonal elements of the density matrix in the slit position representation. As mentioned before, the problem of this representation is the rather arbitrary distinction between diagonal elements and off-diagonal elements. It should be possible to find representations where such a basis-dependent distinction between elements is not needed. 

As mentioned in Sec. \ref{sec:compact}, it would be desirable to satisfy all three conditions given above in order to identify a unique set of elements of reality for the physical system described by the algebra. Any operator expansion that satisfies conditions 2 and 3 cannot satisfy positivity, but we can try to get as close to it as possible. Specifically, we can require that the operators  $\{\hat{\Lambda}(i)\}$ define a quasi-probability with a sum of one, so that the operators must satisfy the completeness relation
\begin{equation}
\sum_i \hat{\Lambda}(i) = \hat{I}.
\end{equation}
This also means that the traces of all reconstruction operators $\hat{R}(i)$ are one. Since orthogonality applies,
\begin{equation}
\hat{R}(i) = \frac{\hat{\Lambda}^\dagger(i)}{\mbox{Tr}(\hat{\Lambda}^\dagger(i))}.
\end{equation}
The normalized reconstruction operators $\hat{R}(i)$ represent the closest approximation to an elementary reality. For the sake of a consistent terminology, these operators represent a quasi-reality. We can now specify the role of quasi-probabilities as statistical weights of the corresponding quasi-realities, where neither the quasi-reality nor the quasi-probability relate to actual observable events. Rather, they are constructed in order to obtain a complete description of causality, highlighting the difference between elements of causality and a mere continuation of reality within the system.   

The idea that quantum statistics could be explained by negative or complex probabilities has a long history going back to the attempts of representing the joint statistics of position and momentum in terms of mathematical functions of the density operator \cite{Wig32,Kir33}. From the mathematical side, a formal extension of probability theories to non-positive values has been considered, e.g. by allowing truth values between $-1$ and $+1$ instead of the exclusive assignment of $0$ and $1$ \cite{Hof09,Bur16,Tof20}. In the present context, it is not necessary to modify the logical concepts associated with probability theories. Instead, it should be remembered that the assignment of joint elements of causality to completely separate measurements and state preparations does not confer a joint reality or truth value to these observables. Quasi-probabilities are convenient expressions of the relation between different possible states and measurements that may represent joint contributions to the statistical weights of two incompatible measurement outcomes as defined by the mathematics of operator expansions. The present analysis shows that the quantum mechanical laws of causality do not require the assumption of joint realities for measurement outcomes that are never observed jointly, which is the reason why negative elements of causality do not imply negative truth values and do not require any modification of our intuitive notion of reality as far as our actual experience is concerned.

Although quasi-probabilities cannot be observed as relative frequencies of actual measurement outcomes, it is often possible to recover the values of quasi-probabilities by combining the information obtained in separate measurements or by subtracting a background noise distribution. The best studied cases are the reconstruction of the Wigner function from marginal distributions \cite{Smi92} and the direct observation of the Dirac distribution using weak measurements \cite{Lun12}. In these cases, the quasi-probabilities represent joint probabilities of observables that cannot be measured jointly, and the marginal distributions must coincide with the actual probabilities of precise measurements performed on only one of the two incompatible observables,
\begin{eqnarray}
\sum_b \hat{\Lambda}(a,b) &=& \mid a \rangle \langle a \mid,
\nonumber \\
\sum_a \hat{\Lambda}(a,b) &=& \mid b \rangle \langle b \mid.
\end{eqnarray}
As explained in \cite{Hof14}, an additional requirement that uniquely identifies the Dirac distribution as the natural joint probability of two non-commuting observables is the condition that the quasi-probability $P(a,b|\psi)$ should be zero for any state $\mid \psi \rangle$ that is orthogonal to either $\mid a \rangle$ or $\mid b \rangle$. Since this requirement can only be satisfied by multiplying the state with a projector, the operator of the quasi-probability is constructed from the product of the two projectors,
\begin{equation}
\label{eq:Dirac}
\hat{\Lambda}(a,b) = \mid b \rangle \langle b \mid a \rangle \langle a \mid.
\end{equation}
As discussed in \cite{Hof11,Hof12a}, this operator expansion not only reproduces many of the features of a classical phase space spanned by $a$ and $b$, it also relates the appearance of complex phases in the quasi-probabilities directly to the action of unitary transformations. 

In the present context, the most important observation is the analogy between phase space statistics and quantum statistics provided by the quasi-probability expansion of the operator algebra. As we have seen above, the positive measurement probabilities of a tomographically complete POVM can only be related to elements of the quantum state through non-positive reconstruction matrices $\hat{R}(i)$. Quasi-probabilities simulate a measurement that is both precise and complete, which would violate the uncertainty principle. The deeper reason for this violation of uncertainty is the necessary appearance of non-positive values in the quasi-probabilities required by the quantum mechanical description of causality. Quasi-probabilities thus reveal the limitations of realist interpretations of the quantum mechanical input-output relations given by Eqs. (\ref{eq:prob}) and (\ref{eq:elements}). 

\section{Entanglement and teleportation}

Entanglement is often considered to be the most characteristic non-classical feature of quantum theory. It is also difficult to explain precisely what entanglement represents. From a statistical viewpoint, it is probably best to say that entangled states represent correlations that exceed the bound expected from the statistics of local quantum states. As we can see from the analysis presented above, the bounds of quantum statistics are related to the non-positivity in any complete description of the operator algebra. It is natural to ask if we can establish a clear and simple relation between the algebra of operator expansions and entanglement. Interestingly, the negativity of the partial transpose of entangled states provides such a relation. Let us consider a maximally entangled state given by
\begin{equation}
\mid E \rangle = \frac{1}{\sqrt{d}}\sum_n \mid n ; n \rangle.
\end{equation}
It is easy to see that the partial transpose of the density matrix of this state can be expressed in terms of the operator $\hat{U}_{\mathrm{SWAP}}$,
\begin{equation}
\left(\mid E \rangle \langle E \mid\right)^{PT} = \frac{1}{d} \hat{U}_{\mathrm{SWAP}}.
\end{equation}
A maximally entangled state can therefore be represented by any complete and orthogonal operator basis. Using the quasi-realities $\hat{R}(i)$, a maximally entangled state can be expressed as
\begin{equation}
\label{eq:E}
\mid E \rangle \langle E \mid= \frac{1}{d} \sum_i
\lambda_i \hat{R}(i) \otimes \hat{R}^*(i),
\end{equation}
where ``$*$'' denotes the complex conjugation of all matrix elements in the $\mid n \rangle$ basis. What is significant about this expression is that it represents the correlations between maximally entangled pairs as coincidences of all $d^2$ quasi-realities $\hat{R}(i)$ with their conjugate $\hat{R}^*(i)$. Importantly, this corresponds exactly to the observed properties of maximally entangled states: for any property $\hat{A}_1$ in system $1$, there is a corresponding property $\hat{A}_2^*$ in system $2$ so that the measurement outcomes of precise measurements of $\hat{A}_1$ and $\hat{A}_2^*$ will be the same. As pointed out by Einstein, Podolsky and Rosen \cite{EPR}, this observation suggests that there is physics beyond the uncertainty principle. The expansion of the entangled state in Eq.(\ref{eq:E}) explains this aspect of quantum physics in terms of correlated quasi-realities. More precisely, Eq.(\ref{eq:E}) shows that the causality between state preparation and spatially separated measurements of systems 1 and 2 needs to be expressed in terms of non-positive elements of causality. The original EPR argument is faulty because it is implied that the ability to make precise predictions based on causality relations requires the pre-existence of elements of reality. However, the uncertainty constraints of quantum measurements ensure that precise predictions based on non-positive elements of causality are possible, because each prediction will only be valid if the corresponding measurement is carried out. This point is further illustrated by Bell's inequality violations, which reveal the non-positive character of the elements of causality hidden by the uncertainty restriction on joint measurements \cite{Hof19}.The elements of causality thus provide a local description of entanglement that explains the violation of positivity bounds in terms of the fundamental relation between different measurements and different kinds of state preparations, neither of which can be reduced to a joint set of positive elements of reality.

All quantum information protocols are based on the extreme precision of causality relations expressed by the non-positive elements $\hat{R}(i)$ which can be accessed if the uncertainties of state preparation and measurement can be compensated in some way. A very good example of a protocol based on the precise causality of maximally entangled states is quantum teleportation \cite{Ben93}. As I have pointed out in previous works, the causality relations of quantum teleportation are entirely local \cite{Hof00,Hof02,Hir13}. In \cite{Hir13}, weak measurement statistics were used to show that quantum teleportation faithfully transfers not just the quantum state, but also the individual fluctuations of the quantum state. For pure states, these fluctuations have no positive representation and therefore correspond to the quasi-realities represented by the elements of causality $\hat{R}(i)$.

Quantum teleportation transfers quantum states by collectively measuring the input $A$ and one part of an entangled pair $R$ on Alice's side in such a way that the measurement is represented by a complete set of maximally entangled states. The outcome obtained in this measurement then describes a unitary transformation that can be applied to the remote system $B$ to recover the original state. The essential piece of information that needs to be transferred from Alice to Bob is the outcome $m$ of the measurement $\{\hat{E}(m)\}$. This outcome contains all of the information necessary to convert the remote system $B$ into the initial state. The physical meaning of the outcome $m$ can be understood in terms of the set of $d^2$ measurement operators $\hat{E}(m)$.  A possible representation of this measurement in terms of the $d^2$ quasi-realities $\hat{R}(i)$ is the permutation representation, where
\begin{equation}
\label{eq:BellM}
\hat{E}(m) = \frac{1}{d^2} \sum_i
\lambda_i \hat{R}_A(i) \otimes \hat{R}_R^*(i+m).
\end{equation}
Note that this representation requires that a shift of the $\hat{R}(i)$ is a unitary operation. Hence all operators $\hat{R}(i)$ must have the same invariant properties. The most likely candidates are the reconstruction operators conjugate to the Dirac distribution operators in Eq.(\ref{eq:Dirac}) for mutually unbiased basis sets. Intuitively, these quasi-realities play the same role as phase space points do in classical physics, and their permutation corresponds to a general reversible transformation in this phase space. The advantage of this representation is that it represents the teleportation measurement as a measurement of the difference $m$ between the quasi-reality $i$ of the input $A$ and the quasi-reality $j=i+m$ of the reference system taken from the entangled pair. Orthogonality then guarantees that each quasi-reality $i$ of the input system $A$ coincides with exactly one quasi-reality $j=i+m$ of the remote system $B$. Mathematically, the result can be obtained by representing the entangled state $\hat{\rho}_{BR}$ using Eq.(\ref{eq:E}), where the complex conjugate elements of the expansion are used to express the expansion of the reference system $R$. The measurement projection then selects the quasi-reality $i+m$ in $B$ for each quasi-reality $i$ in the input $A$. This representation of quantum teleportation corresponds to a classical measurement of the relation between system $A$ and system $R$ when the relation between system $R$ and system $B$ is known. Independent of the statistics of the input state $\hat{\rho}(a)$ we now know the relation between all physical properties in the input $A$ and all physical properties in the remote system $B$. The role of each quasi-reality $i$ is identical to the role of realities described by well-defined values of the physical properties in classical physics. Quantum aspects emerge only because the input must be a positive combination of the quasi-realities $\hat{R}(i)$, and this positivity is only ensured by the constraints on the initial input state $\hat{\rho}(a)$. After the measurement of $m$, the remote state is
\begin{equation}
\hat{\rho}(b|m)=\sum_i \lambda_i \mbox{Tr}(\hat{R}(i) \hat{\rho}(a)) \hat{R}_B(i+m).
\end{equation}
This expression shows that the statistics of the input state $\hat{\rho}(a)$ in A now characterizes the statistics of the conditional state at $B$. However, the physics of the transfer should be explained in terms of the elements of causality $i$, which were available in $B$ before the measurement of $m$. Quantum teleportation merely identifies the accidental relation between elements of causality in $A$ and elements of causality in $B$, exploiting the perfect correlations between these non-positive elements. Positivity is ensured because the input state $\hat{\rho}(a)$  must satisfy positivity and can therefore not be identified with the individual elements of causality. Quantum teleportation thus illustrates the separation of causality from positivity described by the operator formalism. In the operator formalism, causality is not described by a continuation of reality. Instead, the role that elements of reality have in classical physics is taken over by non-positive elements of causality $\hat{R}(i)$ that can represent both continuity and perfect correlations. Since these elements effectively replace the classical realities in the quantum description of causality, it may be justified to call them quasi-realities. However, it should be noted that the classical notion that causality must be mediated by continuous realities is completely unnecessary once it is recognized that state preparation and measurement are always limited by additional constraints that are independent of the internal causality of quantum systems.

\section{Optimal cloning and ideal copies}

The term ``quantum teleportation'' misleading suggests that a physical object is being transferred. However, the physical object onto which the input state is transferred is the same physical object that was initially sent from the entanglement source to Bob. Measurements at $A$ do not cause any physical change of the object $B$. The measurement result $m$ only  updates the available information about $B$.   It is therefore necessary to distinguish between states and physical objects. As the analysis of causality in quantum teleportation shows, the information about the relation of the physical properties of $B$ and the physical properties of $A$ is locally available to Alice because of the perfect correlations between the physical properties of $R$ and $B$ described by the entangled state of the two systems. As shown above, perfect correlations have no positive representation, which means that no local realities can be assigned to the elements of causality. Nevertheless the operator expansion expresses a completely local form of causality. The non-positivity of the operator expansion only indicates that it is physically impossible to isolate elementary realities of a quantum system. This physical impossibility is associated with the uncertainty limits of state preparation and measurements that guarantee the positivity of all observable probabilities. Even pure quantum states describe quantum fluctuations and this means that two systems in the same quantum state are not ideal copies of each other because ideal copies would produce exactly the same measurement outcome whenever the same precise measurement is performed on each of the two copies.  It is therefore a bit strange that ``quantum cloning'' is the name given to the attempted duplication of a state using a single representative of that state.  A true ``clone'' or copy should have the same physical properties and the same measurement outcomes for every precise measurement, and it should be possible to verify this using precise measurements \cite{Hof12b}. Similar to ``teleportation,'' the terminology of quantum cloning reveals a fundamental misunderstanding of the way that the quantum formalism describes the relation between  statistics and physics. It is therefore important to understand how the operator formalism describes the quantum mechanical limit of a copying process.

Physically, the only way to reproduce the same statistics in two systems is to copy each and every property faithfully, since there is no physical record of probabilities in an individual system. Optimal quantum cloning thus represents the closest possible approximation of a universal copy of all physical properties. It is therefore not surprising that the operator expansions discussed above provide us with a particularly simple description of an optimal cloning process. Optimal cloning can be realized by combining the initial state $\hat{\rho}(a)$ with a maximally mixed state of an identical system and selecting the components of the collective two systems state that have a positive exchange symmetry. Since this projector can be expressed by the sum of $\hat{U}_{\mathrm{SWAP}}$ and the identity given in Eq.(\ref{eq:fill}) above, the optimal cloning output can be written as
\begin{eqnarray}
\hat{\rho}_{\mathrm{clone}}(a,a) &=& \frac{1}{2(d+1)} \left( \hat{U}_{\mathrm{SWAP}} + \hat{I} \otimes \hat{I}\right)\left(\hat{I}\otimes \hat{\rho}(a)\right) \left( \hat{U}_{\mathrm{SWAP}} + \hat{I} \otimes \hat{I}\right).
\nonumber \\
 &=& \frac{1}{2(d+1)} \left(\left(\hat{I}\otimes \hat{\rho}(a)\right) + \left(\hat{\rho}(a) \otimes \hat{I} \right) + \hat{U}_{\mathrm{SWAP}} \left(\hat{I}\otimes \hat{\rho}(a)\right) + \left(\hat{I}\otimes \hat{\rho}(a)\right) \hat{U}_{\mathrm{SWAP}}\right).
\end{eqnarray}
As explained in \cite{Hof12b}, the one sided application of $\hat{U}_{\mathrm{SWAP}}$ to the initial product state density matrix represents a component of the state in which all physical properties of the two systems are exactly equal. Effectively, the projection on the positive symmetry of the two systems preferably selects systems that accidentally share all of their physical properties. By using a maximally mixed state as one of the inputs, we can ensure that the probability that all physical properties are accidentally the same is independent of the statistics of the other input state $\hat{\rho}(a)$. The component of the output state $\hat{\rho}_{\mathrm{clone}}(a,a) $that represents the accidental occurrence of perfect copies is given by
\begin{eqnarray}
\hat{C}_{\mathrm{ideal}(a,a)} &=& \frac{1}{2d} \left( \hat{U}_{\mathrm{SWAP}} \left(\hat{I}\otimes \hat{\rho}(a)\right) + \left(\hat{I}\otimes \hat{\rho}(a)\right) \hat{U}_{\mathrm{SWAP}} \right)
\nonumber \\
&=&\frac{1}{2d} \left(\left( \hat{\rho}(a)\otimes \hat{I}\right) \hat{U}_{\mathrm{SWAP}}+ \hat{U}_{\mathrm{SWAP}}(\left( \hat{\rho}(a)\otimes \hat{I}\right)\right). 
\end{eqnarray}
Using any complete operator expansion, $\hat{U}_{\mathrm{SWAP}}$ can be expressed by the sum in Eq.(\ref{eq:swap}). Since the operators $\hat{\Lambda}(i)$ represent quasi-measurements, it is convenient to associate this part of the expansion with the input state $\hat{\rho}(a)$. The ideal pair of copies is then given by
\begin{eqnarray}
\label{eq:Cexpand}
\hat{C}_{\mathrm{ideal}}(a,a) &=& \frac{1}{d} \sum_i \hat{R}(i) \otimes \frac{1}{2}\left(\hat{\Lambda}(i) \hat{\rho}(a) +  \hat{\rho}(a)\hat{\Lambda}(i)\right) \nonumber \\
&=& \frac{1}{d} \sum_i  \frac{1}{2}\left(\hat{\Lambda}(i) \hat{\rho}(a) +  \hat{\rho}(a)\hat{\Lambda}(i)\right) \otimes \hat{R}(i).
\end{eqnarray}
It is easy to verify that the partial trace of either system 1 or system 2 returns the state $\hat{\rho}(a)$ for the other system. However, the operator expansion clearly shows that the ideal pair of copies $\hat{C}_{\mathrm{ideal}}(a,a) $ is not a product state $\hat{\rho}(a)\otimes \hat{\rho}(a)$. instead, the two systems are strongly correlated in their quantum fluctuations. This correlation is necessary because the cloning process is represented by a linear map, which means that the initial quantum state $\hat{\rho}(a)$ can only appear once in the expressions for the ideal cloned pair shown in Eq. (\ref{eq:Cexpand}). It is therefore interesting to consider how the explicit multiplication with $\hat{\rho}(a)$ in system 1 results in the appearance of the same statistics in system 2. Here, the operators $\hat{\Lambda}(i)$ of the operator expansion act like filters that select only the contribution of $i$ in $\hat{\rho}(a)$. In classical statistics, such a filter operation would simply attach a probability of $p(i)$ to a joint reality of $i$ in both systems, corresponding to a copy of the reality $i$ in the initial system onto the other system. For operator expansions, the filter operation is a symmetric operator product of  $\hat{\Lambda}(i)$ and $\hat{\rho}(a)$. In the case of a quasi-probability expansion, the trace of this product represents the quasi-probability of the quasi-reality $i$. However, the operator product introduces additional correlations that cannot be explained by assigning a joint quasi-reality $\hat{R}(i)\otimes \hat{R}(i)$ to the quasi-probability $P(i|a)$. Quantum cloning therefore reveals an interesting limitation of quasi-probability representations, given by the discrepancies $D_i$ between the symmetric operator product with $\hat{\Lambda}_i$ and the contribution of $\hat{R}(i)$ in the expansion,
\begin{equation}
\label{eq:discrepancy}
 D_i(\hat{\rho}) = \frac{1}{2}\left(\hat{\Lambda}(i) \hat{\rho}(a) +  \hat{\rho}(a)\hat{\Lambda}(i)\right) - \mbox{Tr}({\Lambda}(i) \hat{\rho}(a)) \hat{R}(i) \neq 0.
\end{equation} 
Since $\mbox{Tr}(\hat{R}(i))=1$, the trace of this discrepancy is always equal to zero. However, the statistics of a measurement performed on the outputs of quantum cloning requires a product trace with additional measurement operators, and in this product trace, the ordering of the operators in Eq.(\ref{eq:Cexpand}) has a non-trivial effect. It is therefore impossible to represent the ideal copy by a coincidence of identical quasi-probabilities. Instead, the quantum correlations between ideal copies should be characterized by the joint statistics of different measurements performed on systems 1 and 2.

If two different measurements represented by the POVMs $\{\hat{E}_1(b_1)\}$ and $\{\hat{E}_2(b_2)\}$ are performed separately on systems 1 and 2,  the (generally non-positive) measurement probabilities associated with the ideal copy $\hat{C}_{\mathrm{ideal}}(a,a)$ are given by
\begin{eqnarray}
P_{\mathrm{ideal}}(b_1,b_2|a,a) &=& \frac{1}{d} \sum_i \mbox{Tr}(\hat{E}_1(b_1) \hat{R}(i)) \; \mbox{Re}\left(\mbox{Tr}(\hat{\Lambda}(i) \hat{E}_2(b_2) \hat{\rho}(a) \right) 
\nonumber \\
&=& \mbox{Re}\left(\mbox{Tr}(\hat{E}_1(b_1) \hat{E}_2(b_2) \hat{\rho}(a)) \right).
\end{eqnarray}
It should be noted that the first line of this equation includes a product trace of three operators that cannot be separated into a second product trace of $\hat{E}_2(i)$ and $\hat{R}(i)$ and the expansion coefficient $P(i|a)$ of the quasi-probability. This mathematical structure prevents the explanation of quantum cloning in terms of a simple reproduction of the same quasi-reality in the other system. Instead, quantum cloning indicates that all correlations between non-commuting physical properties should be represented by symmetric operator products. As shown in \cite{Hof12a}, the correlations between specific pairs of operators correspond to the Dirac distribution for their eigenstates. However, it is important to recognize that the Dirac distributions of different pairs of physical properties are not consistent with each other, and this inconsistency makes it difficult to transform different Dirac distributions into each other.  It is therefore impossible to identify fundamental quasi-realities $\hat{R}(i)$, even if non-positive probabilities were acceptable. This ambiguity of operator expansions makes it impossible to simplify the deterministic causality described by operator expansions to more direct formulations of universal laws of physics. It is necessary to address this ambiguity by identifying the general role of operator expansions in the description of the physics of cause and effect expressed by Eq.(\ref{eq:prob}), contrasting it with the manner in which elements of reality would perform the same function.

\section{Elements of causality}

The description of ideal copying provides important insights into the way causality in quantum mechanics differs from our classical expectations. An ideal copying process should replicate the causes of all possible effects in such a way that the same elementary cause appears in two separate output systems. However, the ideal copying process described by Eq.(\ref{eq:Cexpand}) does not produce identical pairs of elementary causes $\hat{R}(i)$. Instead, there are additional correlations associated with the discrepancy given by Eq.(\ref{eq:discrepancy}). It is worth considering the role of this discrepancy in more detail. In particular, it is important to note that the density operator $\hat{\rho}(a)$ can be expanded into elements $\hat{R}(i)$, and each of these elements has its own set of discrepancies $D_j$, given by
\begin{equation}
D_j (\hat{R}_i) = \frac{1}{2}\left( \hat{\Lambda}(j) \hat{R}(i) + \hat{R}(i)\hat{\Lambda}(j) \right) - \delta_{i,j} \hat{R}(i).
\end{equation}
The discrepancies for the elements of the operator expansion are thus defined as the deviation of the symmetric product of $\hat{R}(i)$ with $\hat{\Lambda}(j)$ from a projective selection of $\hat{R}(i)$ if and only if $i=j$. Discrepancies of zero indicate that the operators are projectors onto separate subspaces. The fact that the product with $\hat{\Lambda}(j)$ leaves $\hat{R}(i)$ either unchanged or reduces it to zero indicates that the operators $\hat{\Lambda}(j)$ have eigenvalues of one and zero. The requirement that their eigenstates are matched to the support of all operators $\hat{R}(i)$ indicates that the different $\hat{\Lambda}(j)$ project onto orthogonal subspaces. It follows that the only set of operators $\{\hat{\Lambda}(j)\}$ with discrepancies of zero for all combinations of $i$ and $j$ is a positive orthogonal set that satisfies the conditions 1 and 2 given in Sec. \ref{sec:compact}.  As discussed in Sec. \ref{sec:precise}, such a set of operators represents a precise measurement, and the ideal copying process associated with this set of operators would be described by a precise measurement followed by the conditional preparation of the corresponding pure state in the copy. This kind of copying process is indeed possible and easy to realize, but it eliminates all coherences between the eigenstates of the projectors, because the set of operators is incomplete. 

Any complete representation of possible causes must satisfy condition 3 and must therefore violate either condition 1 or condition 2, or both.  More simply put, a complete representation of causality in quantum mechanics requires non-positive elements, either in the description of elementary causes, or in the description of elementary effects. If causes and effects are represented in a time-symmetric manner, condition 2 is satisfied and both causes and effects must be represented by the same set of non-positive operators. Returning to the starting point of Eq.(\ref{eq:prob}), causality must be explained in terms of a complete set of elements since the product trace of any possible combination of states and measurements needs to be explained in a consistent manner. Therefore, condition 3 is indispensable for all elements of causality. In addition, elements of causality should satisfy condition 2 since deterministic causality relations are symmetric in time. The necessary conclusion is that elements of causality cannot satisfy condition 1. They are non-positive and cannot be identified with elements of reality. It should also be noted that elements of causality are not uniquely defined. In fact, the set of operators given in Eq.(\ref{eq:complete}) satisfy conditions 2 and 3, indicating that the elements of any density matrix representation qualify as elements of causality in quantum mechanics. There is no alternative representation that could explain quantum coherences in terms of positive probabilities. 

Elements of causality replace the classical idea of reality by limiting the requirement of positivity to the uncertainty limited descriptions of  state preparation and measurement. This is actually very close to the original justification of quantum mechanics associated with Niels Bohr and Copenhagen. However, it should be emphasized that the physics of state preparation and measurement is still not sufficiently understood because there is no convincing method of resolving the entanglement with the environment that is part of any quantum description of such processes \cite{Zur81,Kar19,Hof20}. The theory presented here describes only the deterministic causality within the quantum system. Both the initialization of the system and its measurement involve interactions with the outside world that cannot be described by internal causalities of the system itself. The formalism of quantum mechanics thus separates the internal causality of a system from the possibility of control from the outside. As the analysis of entanglement suggests, this is achieved  through the selection of specific patterns of uncertainty associated with the means of control outside of the system \cite{Kar19,Hof20}. In the end, external observations of the relations between causes $a$ and effects $b$ are always governed by positive expressions of the form given in Eq. (\ref{eq:prob}). The need for a more complete description of these processes is rooted in the vast number of possible state preparations and measurements. Physics is not concerned with individual events, but with universal laws that govern all possible occurrences.  It is therefore important to note that any universal description of causality necessarily involves non-positive elements to accommodate the full range of possibilities afforded by the optimal control of quantum uncertainties.

\section{Conclusions}

Operator expansions can provide a deeper insight into the difference between quantum statistics and classical statistics. However, these differences cannot be explained by statistics alone. Physics needs to be based on universal rules of causality that cannot depend on the accidental circumstances of individual events. In the discussion above, I have identified the mathematical properties of operator expansions that characterize all causality relations between initial conditions and subsequent observations in quantum physics. The three conditions introduced in Sec. \ref{sec:compact} select operator expansions that are particularly close to the classical notion of statistics as probability distribution over all possible realities of the system. If all three conditions were satisfied, we could identify the associated set of measurement outcomes $\hat{\Lambda}_i$ with the possible realities $\hat{R}_i$ inside the system.  The fact that at most two of them can be satisfied by the operator formalism points towards a fundamental change in the way causality relations work in physics. The key problem is the need for a complete description, and such a complete description necessarily involves non-positive operators. The investigation of entanglement, teleportation and cloning all indicate that the quantum formalism can be understood as a fundamentally non-positive description of causality in which the unavoidable negative values in the elements of causality are hidden by the necessary uncertainties of state preparation and measurement. A universal explanation of causality in quantum statistics thus requires non-positive elements that cannot be isolated in actual experiments due to the uncertainty constraints imposed on the external control of physical systems. Since no consistent description of causality is possible without such elements it may be helpful to think of these non-positive elements of a complete operator expansion as elements of causality, emphasizing the fact that non-positivity is possible because the elements express universal causality relations between the statistics of quantum fluctuations in the initial state and the statistics of quantum fluctuations in the final measurement.
Although a direct observation of these elements of causality is impossible, they are a necessary element in any consistent explanation of the various experimentally observed causality relations, as demonstrated by the observation of ideal correlations in maximal entanglement, quantum teleportation, and optimal quantum cloning. The validity of the operator formalism as a universal description of all possible causality relations in quantum physics therefore indicates that there is no continuation of reality within physical objects. Instead, the external reality of the object is shaped by the various distributions of quantum uncertainties that are required to ensure positive values for all experimentally observable probabilities. Quantum physics therefore describes universal relations between the quantum fluctuations of uncertainty limited state preparations and the quantum fluctuations of the subsequent uncertainty limited observations, where the operator algebra provides the correct mathematical description of the relation between the various uncertainty limited phenomena. 

\section*{Acknowledgments}

This work owes a lot to the very thorough introduction to operator algebra that I  received many years ago from my teacher Prof. G\"unter Mahler at the University of Stuttgart. I would like to dedicate it to his memory.

\end{document}